\documentclass[10pt,aps,prl,reprint,showpacs,superscriptaddress,floats,floatfix,amsmath,amssymb]{revtex4-1}
\usepackage[pdfpagemode=None]{hyperref}
\usepackage{graphicx}
\usepackage{physics}
\usepackage{color}
\begin{document}
\title{How to observe and quantify quantum-discorded states}

\author{{Matthew A. Hunt}}
 \affiliation{School of Physics \& Astronomy, University of Birmingham, B15 2TT, UK}
 \author{Igor V. Lerner}
\affiliation{School of Physics \& Astronomy, University of Birmingham, B15 2TT, UK}
  \author{Igor V. Yurkevich}
\affiliation{School of Engineering \& Applied Science, Aston University, Birmingham B4 7ET, UK}
 \author{Yuval Gefen}
 \affiliation{Department of Condensed Matter Physics, The Weizmann Institute of Science, Rehovot 76100, Israel}


\begin{abstract}
 Quantum correlations between parts of a composite system most clearly reveal  themselves through entanglement. Designing, maintaining, and controlling  entangled systems is very demanding, which raises the stakes for understanding the efficacy of entanglement-free,  yet quantum, correlations,  exemplified by \emph{quantum discord}.
Discord is defined via conditional mutual entropies of parts of a composite system,
and  its direct measurement  is hardly possible even via full tomographic characterization of the system state.  Here we design a simple protocol to detect quantum discord and characterize a discorded state in an \emph{unentangled} bipartite system.
Our protocol is based on an electronic setup and relies on a characteristic of discord that can be extracted from repeated direct measurements of current correlations between subsystems. The proposed protocol opens a way of extending experimental studies of discord  to many-body condensed matter systems.
\end{abstract}

\maketitle

While quantumness of correlations between the parts of a system in a pure state is fully characterized by their entanglement (see Ref.~\onlinecite{Horodecki4} for reviews), mixed states may possess quantum correlations even if they are not entangled. The quantumness of the correlations is properly described in terms of
quantum discord \cite{Zurek:01,Vedral:01}\footnote{For completeness, the full definition of quantum discord\cite{Zurek:01,Vedral:01}  is reproduced in \emph{Supplemental Material}}  which is a discrepancy between  quantum versions of two classically equivalent expressions for mutual entropy in bipartite  systems (see Ref.~\cite{Modi:12,DiscordRev-18,DiscordRev-18b} for reviews).
Any entangled  state of a bipartite system is discorded, but  discorded states may be non-entangled. Although it is entanglement which is usually assumed to be
the key resource for quantum information processes, it was suggested that quantum enhancement of the efficiency of data processing can be achieved in deterministic quantum computation with one pure qubit which uses mixed \emph{separable} (i.e.\ non-entangled) states \cite{Knill:98,Knill-Nat:02,Datta:05,Cable:16}. In such a process, which has been experimentally implemented \cite{White:08}, the nonclassical correlations captured by   quantum discord are responsible for   computational speedup \cite{Datta:08}. Quantum discord was also shown to be the necessary resource for   remote state preparation  \cite{Dakic-NatPhys}, and for the distribution of quantum information to many parties  \cite{Zurek:13,Horodecki:15}. Unlike entanglement, discord is rather robust against decoherence \cite{Mazzola:10}.
Thus, along with   entanglement,   quantum discord can be harnessed for certain types of quantum information processing.

Despite increasing evidence for the relevance of quantum discord, \emph{quantifying} it  in a given quantum state is a challenge. Even full quantum state tomography would not suffice, since determining discord   requires minimizing a conditional mutual entropy over a full set of projective measurements.  An alternative, geometric measure of  discord \cite{Vedral:10,DiscordWitness2,Brodutch:2012,GDDiscordWitn}  has been successfully implemented experimentally \cite{DiscordWitness1,Silva:13,Benedetti:13}. However, geometric discord also faces serious problems. For example,  it can increase, in contrast to the original quantum discord,  even under trivial local
reversible operations on the passive part of the bipartite system \cite{Piani:12} (note, though, the proposal of Ref.~\cite{PhysRevA.88.012120} to mend this deficiency). Most seriously, being a  non-linear function of the density matrix $\rho$, geometric discord can only be quantified via (full or partial) reconstruction of  $\rho$ itself. This severely limits its susceptibility to experiment in the many-body context.

In this Letter we   propose a novel discord quantifier which would  overcome these fundamental difficulties and render quantum discord to be experiment-friendly  for many-body electronic systems, where it has not yet been observed. We present a
  protocol  to detect and characterize quantum discord of any {unknown} mixed state of a generic \emph{non-entangled} bipartite system, implemented in either electronic or photonic setup.  The protocol is based on direct {repeated} measurements of certain two-point correlation functions (which are linear in $\rho$ as any direct quantum-mechanical observable).  While discord cannot be detected by a single linear measurement \cite{Rahimi:10,DiscordRev-18}, we show in detail how \emph{repeated} measurements would allow one to both detect a discorded state and build its reliable quantifier.

Below we will focus on describing how to measure discord in a bipartite two-qubit system. After stating a few  facts about the latter, we will  present an electronic setup, where a bipartite  mixed state can be generated. We then define a relevant two-point correlator, by way of which we can detect and quantify discord.  We demonstrate our protocol by applying it to a few specific states.

 A generic non-entangled bipartite system 
 is described by the   density matrix \cite{Werner:89}
\begin{align}\label{sepstate}
 \rho^{AB}&=\sum_{\nu=1}^Mw_\nu\rho^A_\nu\otimes\rho^B_\nu,
 \end{align}
where  the classical probabilities $w_\nu$ add up to $1$, and each $\rho_\nu^{X} $ describes a pure state of the appropriate subsystem ({$X=A,B$}), so that they can be parameterized as $\rho_\nu^{X }=  \ket{{X}_\nu}\bra{X_\nu}  $.
 It turns out \cite{Vedral:10,Ferraro:10,Modi:12}   that  the mixed state (\ref{sepstate}) is $A$-discorded \footnote{Quantum discord is not necessarily symmetric: one can record discord in one (active) subsystem ($A$) of a bipartite system while the other (passive) subsystem ($B$) might be either discorded or not.} independently of  $\rho_\nu^B$, \emph{unless} $\{ \ket {A _\nu} \} $ form an orthogonal basis. In order to detect and quantify $A$-discord, we propose to utilize this property of state (\ref{sepstate}).
To this end,  we consider correlation functions that are governed by the conditioned density matrix,
 \begin{align}\label{rhoBA}
 \rho^{A|B}=
 \sum_{\nu =1}^nw^B_{\nu}\rho^A_\nu .
\end{align}
 The state described by $\rho^{A|B}$  at the input terminal of  subsystem $A$  evolves  into an out-state described by density matrix $\widetilde{\rho}^{A|B}$ that can  be   diagonalised by adjusting experimentally-controlled parameters of $A$. The coefficients $ w^B_{\nu}$ will depend on the probabilities $w_{\nu}$ and  details of the evolution of subsystem $B$ to be specified later.

We will show that such adjusted parameters are  $ w^B_{\nu}$-independent only if  the states $\{{\ket{{A}_\nu}}\}  $  form an orthogonal basis, i.e.\ $\rho^{AB} $ has no $A$-discord.    Thus their dependence on $ w^B_{\nu}$ is a signature of   $A$-discord. We propose to measure a joint correlation function $K$ of the two subsystems that makes
such a dependence visible,  and to employ $K$ for quantifying  discord. We describe here in detail how to build a reliable discord quantifier based on the correlation function using, for simplicity, a two-qubit bipartite system as an example.

\begin{figure}
\begin{center}
\includegraphics[width=0.47\textwidth]{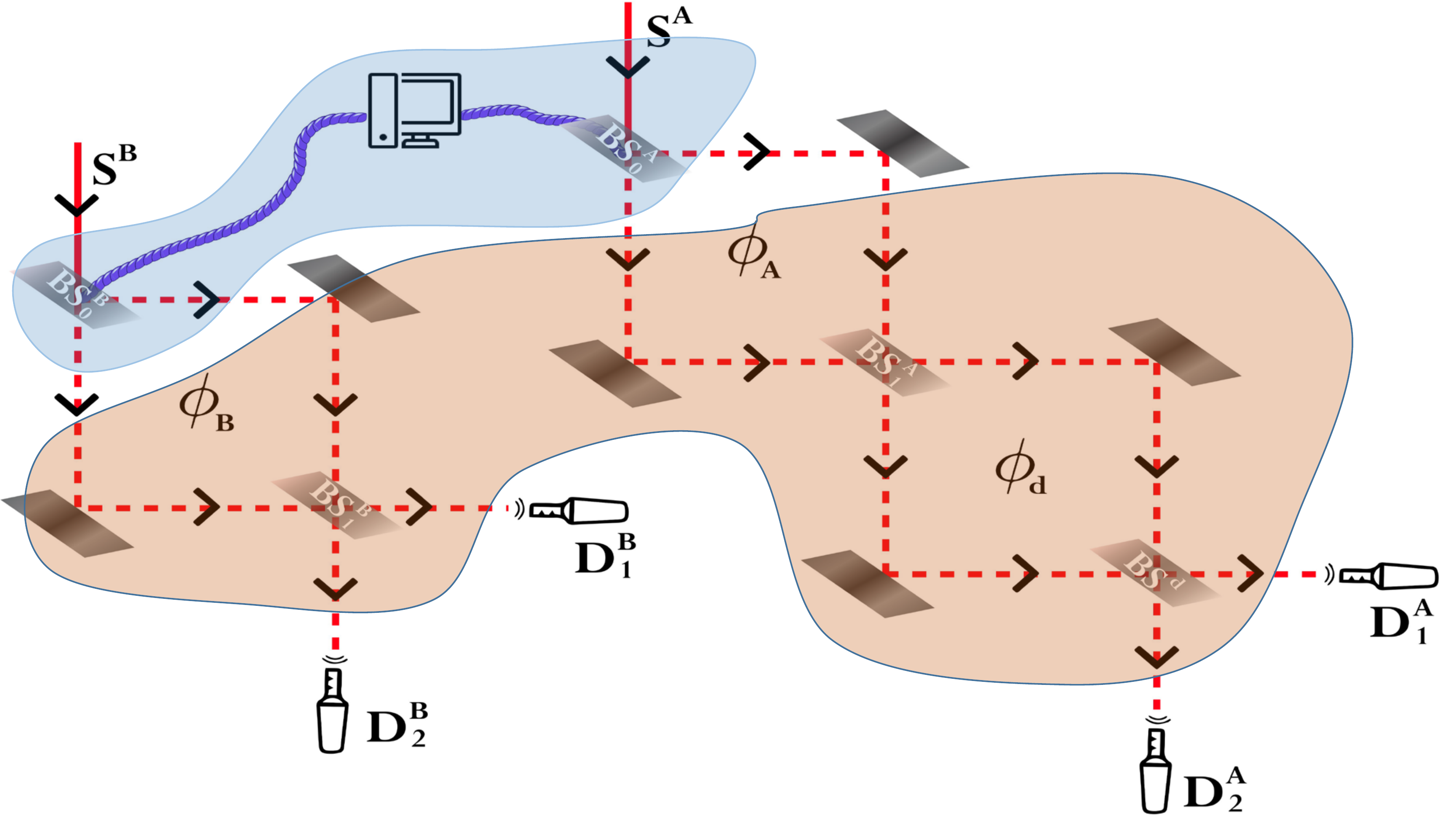}
\end{center}
\caption{\label{MZI} Proposed setup of the bipartite  system made of two Mach-Zehnder interferometers, {MZI$^A$ with the  phase difference $\phi _A$, and MZI$^B$ with  $\phi _B$}. The light blue (light brown) area represents the state-preparation (the state evolution and discord measurement) part of the protocol. Electrons from sources $S^A$ and $S^B$ enter  beam-splitters  BS$_0^{A} $ and BS$_0^B$,    whose random transparencies  are synchronized by a classical computer allowing the creation of mixed states of the form given by Eq. (1). The final state is controlled by transparencies of beam-splitters  BS$_1^{A} $ and BS$_1^B$ and phases   $\phi _{A,B}$, and  is recorded at any pair of detectors $D^A_i$ and $D^B_i$ (with $i=1$ or $2$).  Varying the phase difference $\phi _d$  in the third, detecting  MZI$^d$,  would allow one to identify a state with no $A$-discord as one for which the interference pattern  is suppressed for certain parameters of subsystem $A$ and remains suppressed for any tuning of subsystem $B$ (without adjusting $A$ any further), as illustrated below in Fig.~\eqref{fig2}.}
\end{figure}

It is well known \cite{DiscordRev-18} that a separable state can be prepared by local operations and classical communications. Here we propose a particular way of preparing such a state in a solid state setup. A two-qubit bipartite system with a mixed state, Eq.~\eqref{sepstate}, can be implemented with the help of two Mach-Zehnder interferometers, MZI$^B$ and MZI$^A$, corresponding to subsystems $B$ and $A$ (cf.\ Fig.~\ref{MZI}). Such a system can be realized as   an electron-based setup in a quantum Hall geometry, where the arms of the MZIs are constructed via a careful design of chiral edge modes, and quantum point contacts (QPC) act as effective beam-splitters (BS) \cite{Heiblum:03,Weisz1363,Heiblum:15}. It can also be realized as a photonic device using standard interferometry.

Each qubit is in a quantum superposition of up, $\ket{\uparrow}$,  and down, $\ket{\downarrow}$, states corresponding to a particle  transmitted through the upper or lower arm of the appropriate MZI. The coefficients in each superposition are  determined  by the gate-controlled transparency/reflection of the appropriate BS, with  a phase difference between $\ket{\uparrow} $ and $\ket{\downarrow} $ ($\phi _B$, $\phi _A$) controlled by the  Aharonov--Bohm flux (measured in units of the quantum flux, $hc/e$).

The mixed state, Eq.~\eqref{sepstate},  can be  created  with the help of a \emph{classical} computer  that simultaneously and randomly switches  transparency/reflection   of  BS$^B_0$ and BS$^A_0$  between $n$ values.  The probabilities  $w_ \nu$  in this equation are now  proportional to the time of the pair of BS$_0$ having the appropriate transparencies, provided that the output on the detectors $D^B_{1,2} $ and $D^A_{1,2} $ is averaged over  time intervals much longer than the switching time.

\emph{\textbf{{Principal steps of the proposed protocol.}}} Central to it will be measuring quantum interference in a cross-correlation function between the outputs on the detectors attached to subsystems $A$ and $B$. To this end, we use a third, \emph{detecting} MZI$^d$, cf.\ Fig.~\ref{MZI}. The $\phi _d$-interference pattern vanishes for a set of parameters of subsystem  $A$ for which the density matrix  $\widetilde{\rho}^{A|B}$ (corresponding to the out-state) becomes diagonal in the up-down basis.  This set of parameters of $A$  is independent of the state of $B$ only in the absence of $A$-discord. Its dependence on $B$ will signify the presence of $ A$-discord and allow us to quantify it.

The interference pattern would be revealed in a cross-correlation function of any two operators, $\mathcal{A}$ and $\mathcal{B}$, corresponding to the output observables in subsystems $A$ and $B$. We consider   the joint probability of    particles injected into   $A$ and $B$  to be recorded at the  detectors $D^A_1$ and $D^B_1$:
 \begin{align}\label{K}
 K=\Tr \left[\mathsf{P}_A\,\mathsf{P}_B\, {\mathsf{S}}\,{\rho}^{AB} \,{\mathsf{S}}^{\dagger}\right].
 \end{align}
 Here ${\mathsf{P}_{A,B} }=\ket{\uparrow}\bra{\uparrow}
 $ are the projection operators into detectors {{$D_1^{A,B}$}} in the appropriate space, and the unitary $S$-matrix, \mbox{${\mathsf{S}}= {\mathsf{S}}_B\otimes{\mathsf{S}}_d\,e^{i\phi_d\sigma_3/2}\,{\mathsf{S}}_A$}, describes  independent evolution of the mixed in-state Eq.~\eqref{sepstate} through subsystems $A$,  $B$,   and  the detecting MZI$^d$, see Fig.~\ref{MZI}. The $S$-matrices   for each MZI are products of those corresponding to the beam-splitter and the phase difference accumulated on the opposite arms,
 \begin{align}\label{SB}
 \mathsf{S}_A&=\left(\begin{matrix}
	                        r_A & t_A \\
                            -t^*_A & r _A^*\\
                            \end{matrix}\right)\, e^{\frac{i}{2} \sigma_3 \phi _A} ,
 \end{align}
  and similarly for ${\mathsf{S}}_B$.

Tracing over passive subsystem $B$ reduces the correlation function Eq.~\eqref{K} to
\begin{align}\label{K1}
K({\phi _d})=\Tr_A\,e^{\frac{i}{2}\sigma_3\,\phi_d}\, \widetilde{\rho}^{A|B}\, e^{-\frac{i}{2}\sigma_3\,\phi_d}\,{\mathsf A}\,.
\end{align}
Here $\widetilde{\rho}^{A|B}={\mathsf S}_A\,\rho^{A|B}\,{\mathsf S}_ A^{\dagger}$, and the results of measurements on   passive subsystem $B$ are included into the conditioned density matrix $\rho^{A|B}$ of Eq.~\eqref{rhoBA}  with $w^B_{\nu}=w_{\nu} \Tr_B \big[\mathsf{P} _B{{\mathsf{S}}
_B\rho^{B}_\nu  {\mathsf{S}}_B  ^\dagger
}\big]$, while ${\mathsf A}={\mathsf S}_d^\dagger \,{\mathsf{P}}_A\,{\mathsf S}_d $  with ${\mathsf{S}}^d$ being a scattering matrix through beam-splitter BS$^{d} $  in the detecting MZI$^d$. Choosing it to be a  50:50 BS  makes all  the matrix  elements of ${\mathsf{A}}$  equal to $\frac{1}{2}$.

Due to interference between the $\ket{\uparrow} $ and $\ket{\downarrow} $ states     in  MZI$^d$,   correlation function Eq.~\eqref{K1} oscillates with the phase difference $\phi _d$, controlled in the condensed-matter implementation by  the corresponding Aharonov -- Bohm flux. We parameterize it as
\begin{align}\label{V}
   K({\phi _d})=\mathcal{C}+\left(\mathcal{A}e^{i\phi _d}+{\mathrm{c.c.}}   \right).
\end{align}
This defines the visibility of interference, $\mathcal{V=|A/C|}$, which is the difference between max and min values of $K({\phi _d})$, weighted by its average. It vanishes when $K$  becomes  $\phi _d$-independent. This happens when $\widetilde{\rho}^{A|B}$  in Eq.~\eqref{K1} is diagonal, i.e.\ ${\mathsf{S}}_A={\mathsf{S}}_0$, the diagonalising matrix   for $\rho^{A|B}$.

The final step of the protocol is to check whether ${\mathsf{S}}_0$ is sensitive to changes   in passive subsystem $B$. Such a sensitivity vanishes only if the density matrix of active subsystem $A$ is built on orthogonal states $\{{\ket{A_\nu}}\} $ when discord is absent \cite{YGYL:1}. We will prove the sensitivity to be  a reliable discord witness and show how to build a discord quantifier based on it.

\emph{\textbf{Protocol  implementation.}}  Experimentally, any in-state, Eq.~\eqref{sepstate}, is \emph{repeatedly} generated  in the scheme given in Fig.~\ref{MZI} by  random {simultaneous} changes of transparencies of beam-splitters BS$^B_0$ and BS$^A_0$ with fixed probabilities $w_\nu$.  A set of raw data for the generated in-state should be obtained by varying the phase difference, $\phi _d$, in the detecting MZI$^{d} $ and  measuring the appropriate particle cross-correlation function, Eq.~\eqref{K}.  From this data set, one extracts the visibility, Eq.~\eqref{V}, that is a function of three experimentally controlled parameters, $\alpha $ and $\phi _A$, characterizing the scattering matrix ${\mathsf{S}}_A$, and $\beta $ characterizing  ${\mathsf{S}}_B$ (as the phase difference  $\phi _B$ is always fixed  in the proposed protocol).

\begin{figure*}
\qquad\includegraphics[height=.4\textwidth]{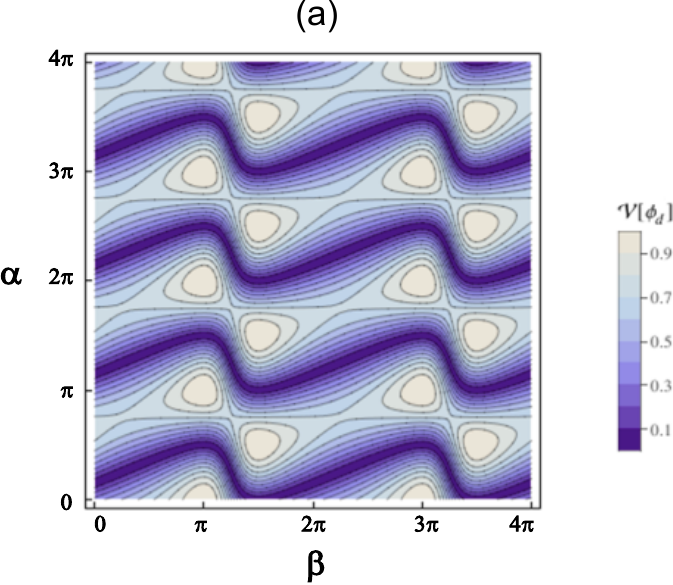}
\quad
\includegraphics[height=.4\textwidth]{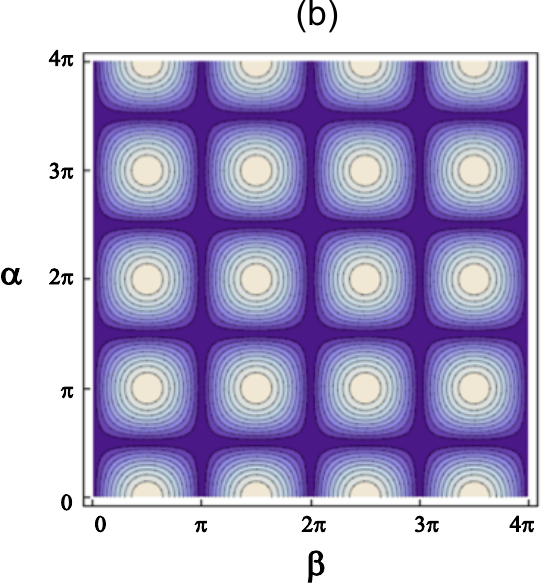}
\vspace*{-6pt}

\caption{\label{fig2}    A striking difference between (a)  discorded and (b)   non-discorded states: \emph{zero-visibility} (dark) lines are sensitive to changes in the state of passive subsystem $B$  (a) and  are independent of this changes in (b).     Here we use the symmetric in-states:   (a) $\rho^{AB} = \frac{1}{2}\left[ \ket{\uparrow\uparrow}\bra{\uparrow\uparrow}{\text+} \frac{1}{2} \ket{\bm {++}}\bra{\bm {++}}\right]   $  and  (b) $\rho^{AB} =\frac{1}{2} \left[\ket{\bm {++}}\bra{\bm {++}}{\text+}\frac{1}{2} \ket{\bm {--}}\bra{\bm {--}}\right]   $ with $\ket{\bm\pm}\equiv \frac{1}{\sqrt{2}}({\ket{\uparrow} \pm\ket{\downarrow} }) $.
Since the states $\ket{+} $ and $\ket{-} $ are  orthogonal whereas $\ket{+} $ and $\ket{\uparrow} $ are not,   these density matrices describe  a  discorded state (a) and a  non-discorded state (b), as explained after Eq.~\eqref{rhoBA}. Any continuous zero-visibility line in (b) can be chosen  for a quantitative characteristic of discord, Eq.~\eqref{W},   cf.\ Fig.~\ref{Discord}.}
\end{figure*}

Fixing also $\beta $, one represents the data as  lines of constant visibility in the $\alpha -\phi _A$ plane, thus producing the visibility landscape.   From this one finds $\phi _{0}$ and $\alpha _0$ that correspond  to zero visibility for this value of $\beta $.  Repeating this for different values of $\beta $, one derives the parametric representation of the zero visibility lines as $\alpha _0({\beta })$ and $\phi _{0  }({\beta }) $.

\begin{figure*}
\qquad\includegraphics[height=.4\textwidth]{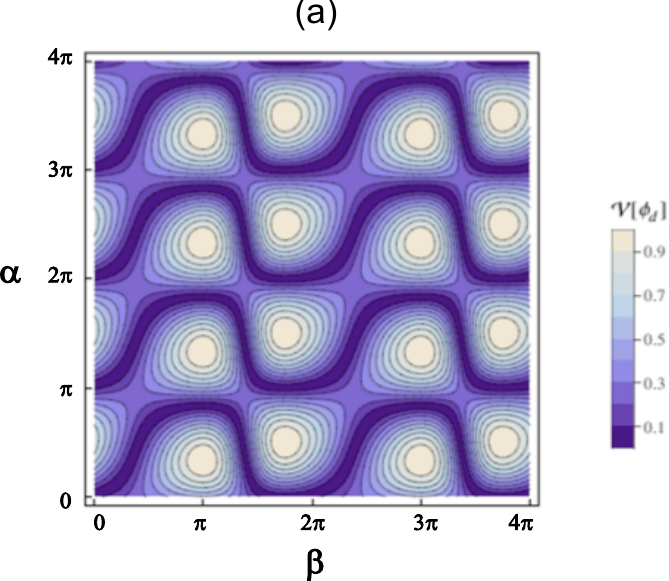}\quad
\includegraphics[height=.4\textwidth]{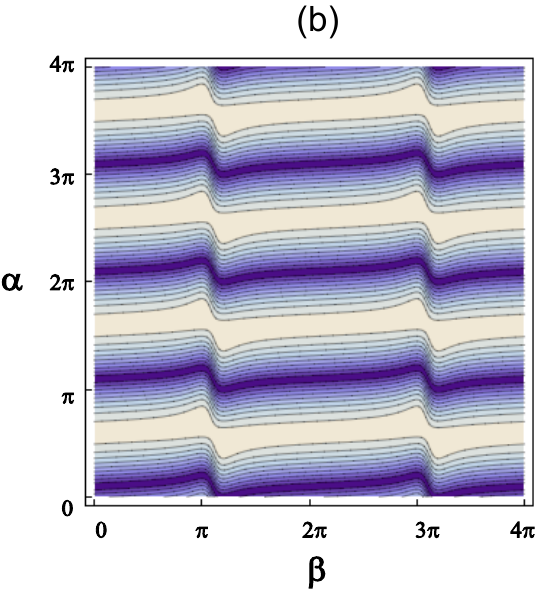}
\vspace*{-6pt}

\caption{ \label{fig3}  The visibility plots for $\rho^{AB}=\frac{1}{2}\ket{\uparrow\uparrow}\ket{\uparrow\uparrow} +\frac{1}{2}\ket{\theta\theta}\bra{\theta\theta}$ where in (a)  $\theta=\frac{5}{6}\pi$ and in (b) $\theta=\frac{1}{5}\pi $. {{These density matrices have similar discord, as can be seen from Fig.~\ref{Discord},  yet their visibility landscapes look completely different: a $\pi $-jump in zero-visibility lines over a small interval of $\beta $ in (a) signifies weak sensitivity with respect to changes in passive system similar to that in (b) featuring the zero-visibility lines with a small non-monotonicity over a large interval.}} }
\end{figure*}

 Let us demonstrate how the protocol works using for illustration simple \emph{real} specified states where the zero-visibility points must have $\phi _0=0\operatorname{mod}(2\pi) $. Hence, $\alpha _0({\beta })$ dependence alone is sufficient for quantifying discord for such states; however,  we   also exploit\cite{YGYL:1} $\alpha ({\phi _A})$ dependence for the example used below that clearly shows $\phi _0=0\operatorname{mod}(2\pi) $, as expected.
 Choosing the in-states defining $\rho^{A}_\nu $ in Eq.~\eqref{sepstate} to be real superpositions of $\ket \uparrow$ and $\ket \downarrow$, we parameterize them as $\ket{A_\nu}{=}\cos \frac{1}{2}\theta_{ \nu}\ket{\uparrow} {+} \sin\frac{1}{2} \theta_{ \nu}\ket{\downarrow} $, and use a similar parameterization for $\rho^{B}_\nu $ with $\theta_\nu\to \widetilde{\theta}_\nu$.  Further choosing these in-states  `symmetric', with   $\theta_{ \nu} =\widetilde{\theta}_\nu$, leads to the parameterization $\rho^{AB}=\sum_{\nu=1}^n w_\nu\ketbra{\theta_{ \nu}  {\theta}_\nu }$,  where we put   $n{=}2$.   Finally, we parameterize the transmission probabilities in  ${\mathsf{S}}_{A,B}$, Eq.~\eqref{SB}, as $|t_A|^2=\sin^2\frac{1}{2}\alpha $ and $|t_B|^2=\sin^2\frac{1}{2}\beta $, making the visibility for a given in-state a function of these two parameters.

In  Fig.~\ref{fig2}, we present the visibility landscape for two different in-states, $A$-discorded and non-discorded, specified in the figure caption.
The dependence  $\alpha _0({\beta })$ corresponding to  the zero-visibility lines in this landscape  reveals a striking difference between the non-discorded and discorded states: the latter shows a strong dependence on $\beta $ while the former is $\beta $-independent; this certainly  works not only for the chosen but for generic mixed states.\cite{YGYL:1}

\begin{figure}
  \includegraphics[width= .8\columnwidth]{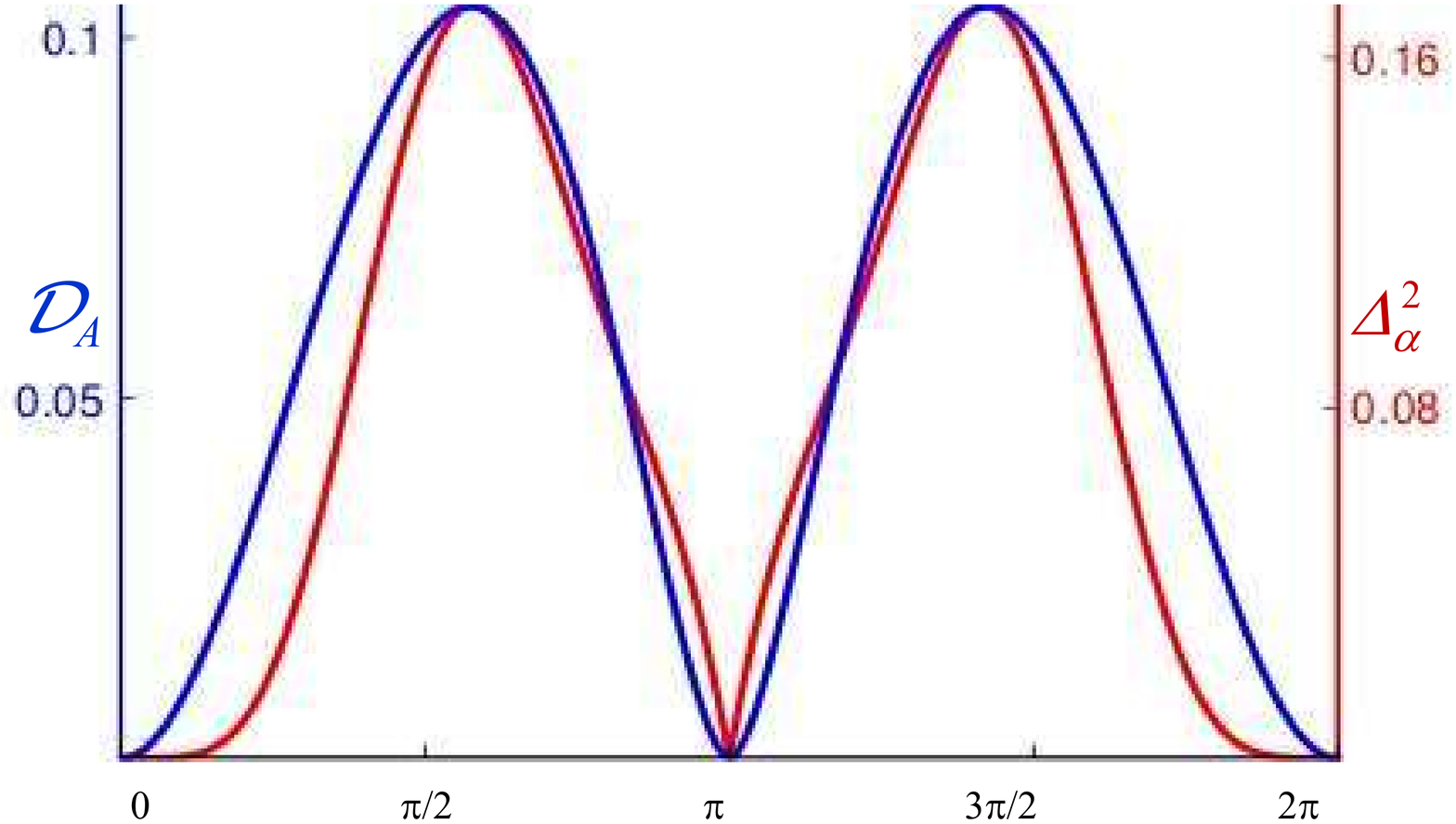}\\
  \caption{The standard definition of discord, $\mathcal{D}_A$, (blue) vs the  alternative quantifier of Eq.~\eqref{W}, $\Delta^2_\alpha $, (red)   for the in-state with the density matrix $\rho^{AB}_\theta=\frac{1}{2}\left[\,\ket{\uparrow\uparrow}\bra{\uparrow\uparrow}+ \ket{\theta\theta}\bra{\theta\theta} \,   \right]$. }\label{Discord}
\end{figure}

{\emph{\textbf{Discord quantifier.}}} The eye-catching signature of discord in Fig.~\ref{fig2}(a) is a high non-monotonicity of the zero-visibility lines, $\alpha _0({\beta })$. By contrast,   such lines are straight for the non-discorded state in Fig.~\ref{fig2}(b).  Note that a  $\pi $-periodic in $\alpha $ pattern of the zero-visibility lines implies that  vertical $\pi $-jumps in zero visibility curves happen for non-discorded states. Hence,   nearly $\pi $-jumps in a zero-visibility curves over a small interval of $\beta $, Fig.~\ref{fig3}(a), signifies weak sensitivity with respect to changes in the passive subsystem similar to that in curves with a small non-monotonicity over a large interval, Fig.~\ref{fig3}(b).  To treat both cases on equal footing,  we  employ the standard deviation of $f_\alpha({\beta})\equiv \cos^2[{\alpha_0({\beta})}]$ from its average over the period as a {quantifier} of such a   sensitivity, which plays the role of a \emph{discord quantifier}:
\begin{align}\label{W}
  \Delta^2_\alpha&=\int\limits_{0}^{2\pi}\frac{\mathrm{d}\beta}{2\pi} \left[f_\alpha({\beta}) -\overline{f}_{\!\alpha}\right]^2\!, &
   \overline{f}_{\!\alpha}&=\int\limits_{0}^{2\pi} \frac{\mathrm{d}\beta}{2\pi}f_\alpha({\beta}).
\end{align}

 This quantifier gives similar results for the two sets of symmetric in-states in Fig.~\ref{fig3}. Both have the density matrix  \mbox{$\rho^{AB}_\theta=\frac{1}{2}\left[\ket{\uparrow\uparrow}\bra{\uparrow\uparrow}+ \ket{\theta\theta}\bra{\theta\theta}    \right] $} with different $\theta$. For $\theta\!=\!0 $, $ \rho^{AB} =\ket{\uparrow\uparrow}\bra{\uparrow\uparrow}$ is a pure state  with no discord, and likewise discord is  absent for  $\theta=\pi $ when $\rho^{AB}_\theta\to\frac{1}{2}\left[\,\ket{\uparrow\uparrow}\bra{\uparrow\uparrow}+ \ket{\downarrow\downarrow}\bra{\downarrow\downarrow} \,  \right]$. Thus, discord is small for $\rho_\theta^{AB} $ with $\theta$ approaching either $0$ or $\pi $, cf.\ Fig.~\ref{Discord}.

This suggested quantifier is convenient and, although it is  by no means unique, it works remarkably well:  its similarity to  quantum discord in its original definition  is    quite  appealing, as illustrated for $\rho_\theta^{AB} $ of the above example  in Fig~\ref{Discord}.    It is straightforward to prove\cite{YGYL:1}  that this measure is reliable:   it  vanishes for any non-discorded state and does not change with a unitary transformation on passive subsystem $B$.

\emph{\textbf{ Conclusion. }}
 We   have proposed  a new characterization of quantum discord based on measuring cross-correlations in non-entangled bipartite systems and thus linear in density matrix $\rho$,  in contrast to other quantifiers, notably geometric discord, that require full or partial quantum tomography for reconstruction of $\rho$. The linearity of the proposed quantifier opens a path to extending experimental research of discord into electronic condensed matter systems. We have considered in detail one possible implementation via devices built of Mach -- Zehnder interferometers in quantum Hall systems, where our quantifier
 is quite robust against external noise and fluctuations: as long as the Aharonov--Bohm oscillations are resolvable \cite{Heiblum:03}, the appropriate interference pattern may serve as a pictorial discord witness, as illustrated above in Figs.~\ref{fig2} and \ref{fig3}. Finally, our discord quantifier  is qualitatively consistent, and quantitatively very close  to the original measure.

 The relative simplicity of this protocol, and the fact that it is based on presently existing measurement technologies and available setups (electronic Mach-Zehnder interferometers) is bound to stimulate experiments in this direction. While the present
analysis addresses discord of bi-partite systems, an intriguing generalization of our protocol to multiply-partite systems is possible by introducing a number of coupled interferometers
 Extension of our protocol to anyon-based states (employing anyonic interferometers) or other topological states may open the horizon to topology-based study of discord.

\emph{\textbf{ Acknowledgements. }}This work was supported by the Leverhulme Trust Grants RPG-2016-044 (IVY), VP1-2015-005 (IVY, YG), the Italia-Israel
project QUANTRA (YG), and the DFG within the network CRC TR 183, C01
(YG). The authors (IVL, IVY and YG) are grateful for hospitality extended to them at the final stage of this work at the Center for Theoretical Physics of Complex Systems, Daejeon, South Korea.

%

\newpage
\onecolumngrid
\appendix
\begin{center}
    \large Supplemental Online Materials
\end{center}

\renewcommand{\theequation}{S.\arabic{equation}}
\setcounter{equation}{0}

\subsection{Quantum Discord}
Quantum Discord [1,2] exemplifies the difference between classical and quantum correlations of two (sub)systems ({$A$ and $B$}), as quantified by mutual information. The latter, which is a classical measure of correlations between  {$A$ and $B$}, is defined   as ${I}(A:B)\equiv H(A)+H(B)-H(AB)$, where    the Shannon entropy $H(A)\equiv-\sum_a p_{a}\log p_{a}$ with $a$ being the possible values that a classical variable $A$ can take with the probability $p_a$, while the joint entropy $H(AB) $ is that of the entire system $A\bigcup B$. An alternative way of writing a classically equivalent expression to $I(A:B)$ is $J(A:B)\equiv H(A)-H(A|B)$, with $H(A|B)\equiv H(AB) -H(B)$ being the conditional entropy which is the uncertainty remaining about $A$ given a knowledge of $B$'s distribution.

The quantum analogues to these expressions can be obtained [1-4] by replacing the Shannon entropies for the probability distributions with the corresponding von Neumann entropies for QM density matrices, $S(\rho )=-\Tr\{\rho \log\rho \}$. The quantum analogue of $I(A:B)$ is then straightforward to define,
\begin{align}
\mathcal{I}_{\mathrm{Q}}(\rho^{AB})&\equiv S(\rho^A)+S(\rho^B)-S(\rho^{AB}), \label{I(rhoAB)}
\end{align}
where $\rho^A,$ $\rho^B$ are the reduced density matrices on either subsystem. However, the straightforward analogue to the classical conditional entropy is not that useful since if we defined $S({B|A})$ as $S({A,B})-S({A})$, this quantity would be negative, e.g., in the case when   subsystems $A$ and $B$ are in a pure state. Instead, the quantum conditional entropy $S({B|A})$ is defined as the average von Neumann entropy of states of $B$ after a measurement is made on $A$.

The result of a `measurement' will depend on the basis we pick for our measurement projectors. Post-measurement density matrix becomes
\begin{align}
\rho^{AB}\to\sum_{\mu}\,p_{\mu}^A\,\Pi_{\mu}^A\otimes\rho_{B|\Pi^A_{\mu}}
\end{align}
where $\rho_{B|\Pi^A_{\mu}}$ is the density matrix conditional on some measurement on $A$ as follows,
\begin{align}
\rho_{B|\Pi^A_{\mu}}\equiv\frac{1}{p_{\mu}^A}\,
\Tr_{A}\left(\Pi^A_{\mu}\otimes\Pi^B\right)\rho^{AB}\left(\Pi^A_{\mu}
\otimes{\Pi}^B
\right)\,
\qquad p_{\mu}^A=\Tr \left(\Pi^A_{\mu}\otimes\Pi^B\right) \rho^{AB}
\label{RhoCon}
\end{align}

Using this conditional state, Eq.~(\ref{RhoCon}), we may extract the entropy $S(\rho_{B|\Pi^A_{\mu}})$ which gives us the amount uncertainty of the state of $B$ given this projection onto $A$. We may then obtain the conditional entropy after a complete set of measurements on $A$, $\{\Pi^A_{\mu}\}$,
\begin{align}
S(B|\{\Pi^A_{\mu}\})\equiv\sum_{\mu} p^A_{\mu} S(\rho_{B|\Pi^A_{\mu}})
\end{align}
From which a generalisation of $J(A:B)$ can be constructed,
\begin{align}
\mathcal{J}_A(\rho^{AB})\equiv S(\rho^B)-S(B|\{\Pi^A_{\mu}\}),
\end{align}
where one final ingredient has also been added in order to remove the dependence on the measurement basis; we maximised over all complete measurement bases, essentially we pick the \emph{best} measurement basis (that is the one where we are able to reduce our ignorance about subsystem $B$ the most).

Having defined two quantities which would be classically equivalent, the difference between the two could be thought of as a measure of `quantumness'. This quantity was termed the \emph{quantum discord},
\begin{align}
\mathcal{D}_A(\rho^{AB})\equiv\min_{\{\Pi^A_{\mu}\}}\left[\mathcal{I}(\rho^{AB})-\mathcal{J}_A(\rho^{AB})\right]=\min_{\{\Pi^A_{\mu}\}}S(B|\{\Pi^A_{\mu}\})-\left[S(\rho^{AB})-S(\rho^A)\right].
\end{align}
Note that since $\mathcal{J}$ is not symmetric about which subsystem we perform the measurement on, neither is discord and in general $\mathcal{D}_A(\rho^{AB})\neq \mathcal{D}_B(\rho^{AB})$.

\subsection{Derivation of the discord quantifier}

In what follows we  simulate the visibility data for  specified in-states. We will find expressions for zero-visibility lines and show how to construct the visibility landscapes for for any specified non-entangled state. In particular, we illustrate how to do this for the in-states employed in  Figs.~\ref{fig2} and \ref{fig3}; we also demonstrate that  in this case $\phi _0=0\operatorname{mod}(2\pi) $, as expected, \ref{fig5}.  However, we stress that the known states used for illustrative purposes only:  the protocol as described in this section can be experimentally implemented for any non-entangled state.

\emph{\textbf{ {Visibility for specified  in-states.}}}
We have defined the visibility $\mathcal{V} $  in terms of the parameters defining correlation function $K$, Eq.~\eqref{V}.
Let us   parametrise $\rho_\nu^A$ in Eq.~\eqref{K1} via the unit vector  on the appropriate Bloch sphere as $\rho_\nu^A=\frac{1}{2}({1+{\bm{n}_{\nu}\cdot{\bm{\sigma}}}})\equiv \ketbra {\bm n_\nu}$ with
\begin{align}
\label{n-nu} \begin{aligned}
    {\bm n}_{\nu}&=(\sin\theta_{\nu}\cos\phi_{\nu},\sin\theta_{\nu}\sin\phi_{\nu},\cos\theta_{\nu})\,,
\\
\ket{{\bm n}_{\nu}}&=\cos(\tfrac{1}{2}\theta_{\nu} )\ket{\uparrow}+e^{i\phi_{\nu}}\sin(\tfrac{1}{2}\theta_{\nu} )\ket{\downarrow}
\end{aligned}
\end{align}
 Substituting this into Eq.~\eqref{rhoBA} we obtain the conditioned density matrix in Eq.~\eqref{K1} as
 \begin{align}\notag
\rho^{\rm A|B}&=\tfrac{1}{2}\left[W_B\mathrm{I} +C {\bm n}\cdot{\bm\sigma}\, \right]\\&=\tfrac{1}{2}\big[(W_B-C)\mathrm{I}+C\ketbra{{\bm n}} \big] .\label{K3}
\end{align}
Here $W_B=\sum_{\nu} w_\nu^B$, with $w_\nu^B$ defined after Eq.~\eqref{K1}, and the unit vector ${\bm{n}}{=} (\sin\vartheta\cos\varphi,\sin\vartheta\sin\varphi,\cos\vartheta)$ with the normalisation constant $C$ given by
\begin{align}\label{Cn}
 C{\bm{n}}&=\sum_{\nu} w_{\nu}^B{\bm n}_{\nu} .
\end{align}
State $\ket {\bm n}$ is parameterized via $\vartheta$ and $\varphi $ as in Eq.~\eqref{n-nu}:
\begin{align}\label{ketn}
    \ket{{\bm n} }=\cos \tfrac{1}{2}\vartheta  \ket{\uparrow}+e^{i\varphi }\sin \tfrac{1}{2}\vartheta  \ket{\downarrow}.
\end{align}
From Eq.~\eqref{K3} and Eq.~\eqref{ketn} follows that matrix ${\mathsf{S}}_0$ that diagonalises $\rho^{\rm A|B}$ and thus defines zero-visibility lines obeys, up to a phase factor, ${\mathsf{S}}_0\ket{\bm n}=\ket \uparrow$ or $\ket \downarrow$. Using the parameterization of Eq.~\eqref{SB} for ${\mathsf{S}}_A\to {\mathsf{S}}_0$, i.e.\ $|r_0|= |\cos \frac{1}{2}\alpha _0| $ and $\phi_A\to \phi _0$, we find that the zero-visibility lines are given by
\begin{align}
\cos ^2  \alpha_0& =  n_z^2, \,\text{ and }\, \tan \Phi _0=-n_y/n_x,
\end{align}
with $\Phi _0=\phi_0+\arg({r_0/t_0})$.
More generally,   one expresses  constant visibility lines,  $\mathcal{V=|A/C|}=\mathrm{const}$, via the coefficients in Eq.~\eqref{V} obtained   from Eq.~\eqref{K3}  and Eq.~\eqref{ketn} as $\mathcal{C}=\frac{1}{2}W_B$ and $\mathcal{A}=\frac{1}{2}\big(c_\uparrow c^*_\downarrow\big)$, with
$
c_{\uparrow} =r_A\cos\frac{\vartheta}{2}+t_A\sin\frac{\vartheta}{2}e^{i(\varphi-\phi_A)}\,,$ and $
c_{\downarrow}=-t^*_A\cos\frac{\vartheta}{2}+r^*_A\sin\frac{\vartheta}{2}e^{i(\varphi-\phi_A)}
$.
For the real in-state used in the example of Fig.~\ref{fig2}, we have $\mathcal{C}=\frac{1}{2}\big[1+\sum_{\nu} w_\nu \cos({\beta -\theta_\nu})\big]$ and $\mathcal{A}=\frac{1}{4}\sin ({ \alpha -\vartheta})$,  where $\vartheta$ is expressible via $\theta_\nu$ with the help of Eq.~\eqref{Cn}.

 \emph{\textbf{{Zero-visibility lines for non-discorded states.}}} We now show in detail that the zero-visibility condition is independent of the measurement on $B$ \emph{if and only if} the mixed state $\rho^{AB} $ is non-discorded. The dependence of the measurement on $B$ enters the zero-visibility equation via the coefficients $ b_\nu $. This dependence vanishes if all vectors ${\bm n}_{\nu}$ become parallel to each other, in which case vector ${\bm n}$, Eq.~\eqref{Cn}, does no longer depend on coefficients $w^B_{\nu}$ (i.e. sub-system B). This is equivalent to the statement that all states $\ket{{\bm n}}_{\nu}$ are either coincide (up to a phase) or orthogonal to each other. In general, states are separated into two mutually orthogonal groups.
	
As the mutual orthogonality of the in-states $\{{\ket{A_\nu} }\} $ is a necessary and sufficient condition [4-6]  for the mixed state described by the density matrix $\rho^{AB} $, Eq.~\eqref{sepstate}, to be $A$-discorded, we have proved that the absence of the $\beta$-dependence in the zero-visibility lines signifies the absence of $A$-discord. Graphically, this leads to horizontal and vertical zero-visibility lines like those in Fig.~\ref{fig2}(b). On the contrary,  curving zero-visibility lines, as in Figs.~\ref{fig2}(a) and \ref{fig3}, give a striking, experimentally accessible signature of   quantum discord.

 In the above example  of applying the protocol, Fig.~\ref{fig2}, we have  employed the real in-state with $\theta_{\!B_\nu}=\theta_{\!A_\nu}\equiv \theta_\nu  $ described by the density matrix  \mbox{$\rho^{AB} =\frac{1}{2} \ket{\theta_1\theta_1}\bra{\theta_1\theta_1}+ \frac{1}{2} \ket{\theta_2\theta_2}\bra{\theta_2\theta_2}  $}. In this case (and for any real in-state), $n_{1y}=n_{2y}=0  $ in Eq.~\eqref{n-nu}, so that $\phi _{B_0}{=}0$ or $\pi$.  This is clearly seen from  the visibility landscape in  Fig.~\ref{fig2}. This landscape is  constructed from Eqs.~\eqref{n-nu}   for the two in-states, one with $\theta_1{=}\pi/2$ and $\theta_2{=}3\pi/2$, and the other with $\theta_1{=}0$ and $\theta_2{=}\pi/2$, both for the fixed $\alpha{=}\pi/3$. Since  the inner product $\braket{\theta_1,\theta_2} $ is zero for the first case and non-zero (and $\ne1$) for the second, these states are, respectively, non-discorded and discorded. This is demonstrated in Fig.~\ref{fig5} where    the visibility landscape on the $\beta$-$\alpha$ plane shows a striking $\alpha$-dependence which is a signature of $A$-discord. The application of the protocol is illustrated by further examples in \emph{Supplemental Material}  where   the visibility landscape is drawn for a set of further examples,  representing the in-states which are asymmetric ($\theta_{\!B_\nu}\ne\theta_{\!A_\nu}$), have different probabilities $w_\nu$, non-zero phases, and more than two constituents ({$n\ne2$}).
\subsection{Zero Discord: Grid or Barcode}\label{sec:GridorBarcode}
We demonstrated above  that a lack of dependence on $\beta$ of the lines of zero visibility (i.e the lines are straight as a function of $\beta$) means zero discord, but often in cases of zero discord we also see the emergence of vertical straight lines (zero's of visibility  which are $\alpha$ independent). We find that the two scenarios of just horizontal lines (barcode) or a grid-like scenario (where both vertical and horizontal zero visibility lines are present) whilst both referring to zero discord states refer to two different routes of getting there. Horizontal lines mean the two subsystems are entirely uncorrelated, whilst grid-like means the two subsystems are correlated but only classically.
\begin{figure*}
\includegraphics[height=.45\textwidth]{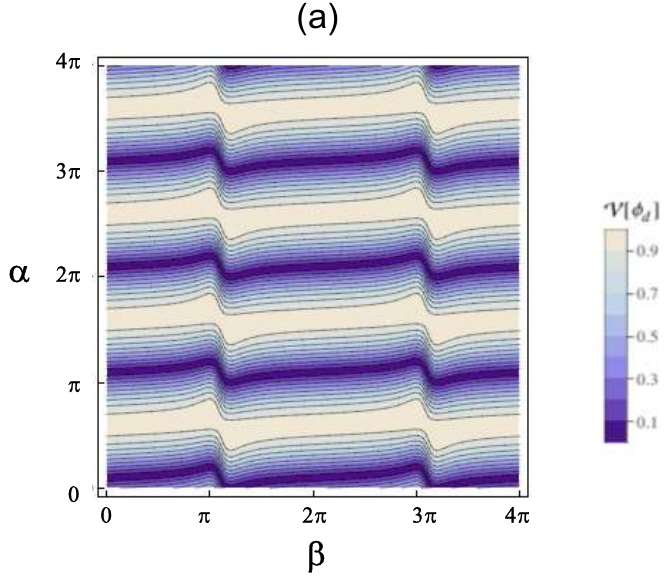}\quad
\includegraphics[height=.45\textwidth]{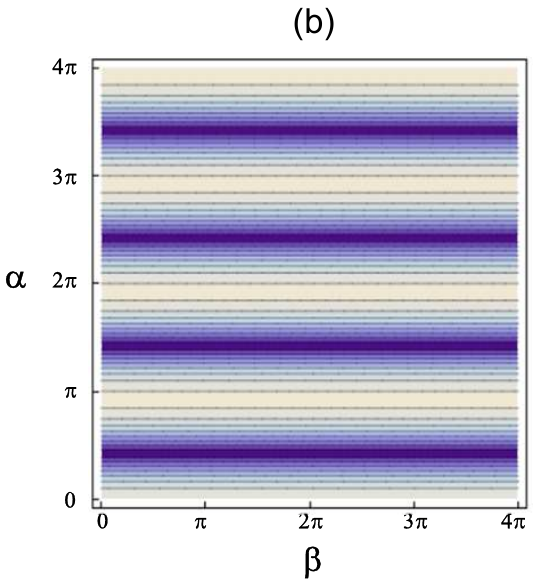}
\vspace*{-6pt}

\caption{ \label{gridorbarcode} The visibility plots for  (a) $\rho^{AB}=\frac{1}{2}\ket{00}\bra{00}+\frac{1}{2}\ket{+0}\bra{+0}$, with $\phi_A=0$   and  (b) $\rho^{AB} =\frac{1}{5}\ket{--}\bra{--}+\frac{4}{5}\ket{++}\bra{++}$. (a) display the 'grid-like' visibility characteristic of a density matrix which is correlated between $A$ and $B$ subsystems, but undiscorded i.e the correlations are classical only. The barcode graph of (b) is a result of a density matrix which is completely uncorrelated between $A$ and $B$ subsystems. $\phi_A$ has been fixed as $\phi_{A_0}=0$ in both plots.}
\end{figure*}

\textbf{\textit{Grid-like}.}
A gridded graph such as Fig. \ref{gridorbarcode}(a), as well as one of Fig. \ref{fig:AlphaVsBetaQQ}(a), is produced when the state is \emph{classically correlated}, that is when no information about the correlations between subsystems A and B is lost when one makes the correct choice of measurement on subsystem-B.

A classically correlated state (with respect to measurement on A), means that the state described by $\rho^A_1$ is orthogonal to that described by $\rho^A_2$. States of this form may always be reduced to a mixed state of the form $\sum_{i=1}^2w_{\nu}\rho^A_{\nu}\otimes\rho^B_{\nu}$, where $\rho^A_{\nu}$ are pure but $\rho^B_{\nu}$ are, in general, mixed and not equal ($\rho^B_1\neq \rho^B_2$). If $\rho^B_1= \rho^B_2$, $\rho^{AB}$ is uncorrelated and we obtain the barcode images in the next part of this section.

The complex amplitude of the oscillatory part of the correlation function was defined in the main text,
\begin{align}
{\cal A}=\left[\mathsf{S}_A\,\rho^{A|B}\,\mathsf{S}_A^{\dagger}\right]_{12}\,A_{21}\,,
\end{align}
where
\begin{align}
\rho^{A|B}=\sum_{\nu}\,w_{\nu}^B\,\rho_{\nu}^A\,,\quad \mathsf{A}=\mathsf{S}_d^{\dagger}\,\mathsf{P}_d\,\mathsf{S}_d\,,\quad
w_{\nu}^B=w_{\nu}\,{\mathrm Tr}_B\,\rho_{\nu}^B \mathsf{S}_B^{\dagger}\,\mathsf{P}_B\,\mathsf{S}_B\,.
\end{align}
Assuming the simplest model of the detector QPC,
\begin{align}
\mathsf{S}_d=\frac{1}{2}\,\left(\begin{array}{cc}
1 & 1 \\ -1 & 1
\end{array}\right)\,,\quad
\mathsf{A}=\frac{1}{4}\,\left(\begin{array}{cc}
1 & 1 \\ 1 & 1
\end{array}\right)\,,
\end{align}
and parametrising A-states as
\begin{align}
\rho_{\nu}^A=\frac{1}{2}\,\left[1+{\bf n}_{\nu}\,{\bm\sigma}\right]
\end{align}
the complex amplitude is written as:
\begin{align}\label{N}
{\cal A}=\frac{1}{8}\,{\bf N}\,\left[\mathsf{S}_A\,{\bm\sigma}\,\mathsf{S}_A^{\dagger}\right]_{12}\,,\quad {\bf N}=\sum_{\nu}w_{\nu}^B\,{\bf n}_{\nu}\,.
\end{align}
The parametrisation of $\mathsf{S}_A$ as
\begin{align}
r_A=e^{i\varphi_r}\,\cos\frac{\alpha}{2}\,,\quad t_A=e^{i\varphi_t}\,\sin\frac{\alpha}{2}\,,\quad \varphi_{\pm}=\varphi_r\pm\varphi_t\,,
\end{align}
with the use of
\begin{align}
\left[\mathsf{S}_A\,{\bm\sigma}\,\mathsf{S}_A^{\dagger}\right]_{12}=\left(r_A^2-t_A^2,\,-i(r_A^2+t_A^2),\,-2r_At_A\right)\,,
\end{align}
leads to another representation
\begin{align}
{\cal A}=e^{i\varphi_+}\frac{1}{8}{\bf N}{\bf a}\,,\quad {\bf a}={\bf a}_1+i{\bf a}_2\,.
\end{align}
The complex vector ${\bf a}$ is a linear combination of two orthonormal vectors
\begin{align}
{\bf a}_1&=\left(\cos\alpha\,\cos\varphi_-\,,\cos\alpha\,\sin\varphi_-\,,-\sin\alpha\right)\,,\\
{\bf a}_2&=\left(\sin\varphi_-\,,-\cos\varphi_-\,,0\right)\,,
\end{align}
The only dependence on B-system may emerge in the vector ${\bf N}$ through the coefficients $w_{\nu}^B$ in the linear combination. \\
Writing all vectors in spherical system of coordinates
\begin{align}\label{n}
{\bf n}=&\left(\sin\theta\cos\phi\,,\sin\theta\sin\phi\,,\cos\theta\right)\,,\quad {\bf N}=N\,{\bf n}\,,\\
{\bf n}_{\nu}=&\left(\sin\theta_{\nu}\cos\phi_{\nu}\,,\sin\theta_{\nu}\sin\phi_{\nu}\,,\cos\theta_{\nu}\right)\,,
\end{align}
\\
In a general situation, the condition of vanishing oscillations is the orthogonality of vector ${\bf N}$ to the set of orthonormal vectors ${\bf a}_1$ and ${\bf a}_2$. It means that vector ${\bf N}$ is parallel to
\begin{align}
{\bf a}_3=\left[{\bf a}_2\times{\bf a}_1\right]=\left(\sin\alpha\cos\varphi_-\,,\sin\alpha\sin\varphi_-\,,\cos\alpha\right)\,.
\end{align}
This statement can be written as
\begin{align}\label{solution}
{\bf n}=\pm{\bf a}_3\,,
\end{align}	
or, cf Eq.~(\ref{n}), in the following form:
\begin{align}\
\alpha=&\theta\quad \mathrm{mod} \,\,\pi\,,\\
\varphi_-=&\phi\quad \mathrm{mod} \,\,2\pi\,.
\end{align}
The angles $\theta$ and $\phi$ may depend on B-system through the coefficients $w_{\nu}^B$ in the linear combination, Eq.~(\ref{N}).\\
\\
{\bf NB} The solution Eq.~(\ref{solution}) can be written only for non-zero vectors ${\bf N}$, when a unit vector ${\bf n}$ is well defined. There is an extra solution for zero visibility lines when
\begin{align}
{\bf N}=\sum_{\nu}\,w_{\nu}^B{\bf n}_{\nu}=0\,.
\end{align}
This equation can be satisfied only when all vectors ${\bf nu}$ are parallel to each other, i.e. when system A happens to be classical. Then there might be solutions of
\begin{align}
\sum_{\nu}(\pm)w_{\nu}^B=0\,,
\end{align}
that define values of parameters (describing system B only) where oscillations vanish.

\textbf{\textit{Example: the states with $\phi_{\nu}=0$.}}
This corresponds to the configuration when all vectors, ${\bf n}_{\nu}$ and, therefore, ${\bf N}$, lie in $(x,z)$-plane. It is sufficient to consider $\mathsf{S}_A$ rotating in that plane only which corresponds to the real $\varphi_{\pm}=0$. The amplitude of oscillations becomes
\begin{align}
{\cal A}\sim {\bf N}{\bf a}_1=N\,\sin(\theta-\alpha)\,.
\end{align}
Generic zeros $\alpha=\theta$ mod $\pi$ and extra solutions for classical A-system are found from $w_{1}^B=w_{2}^B$ (for two terms case).

\textbf{\textit{Grid-like}.}
In Fig. (\ref{gridorbarcode}a) the pattern is grid-like which can be observed only for zero-discord system with orthogonal states $\rho_{\nu}^A$ under the extra condition $w_1^B=w_2^B$. For the example used, $w_1^B=(1/10)(1+\sin\beta)$ and $w_2^B=(4/10)(1-\sin\beta)$, leading to vertical zero visibility lines at $\sin\beta=3/5$.

\textbf{\textit{Barcode-like}.}
Zero-visibility lines like those shown in Fig. (\ref{gridorbarcode}b) again fall into the zero discord category, but without the vertical lines because the states $\rho_{\nu}^A$ are not mutually orthogonal.

\subsection{Further examples of the protocol}
We now further demonstrate how our protocol could be performed in practise. We provide different examples to those given in the main text.

\textbf{\textit{States with no phase differences}.}
This family of states is described by the following density matrix:
\begin{align}
\rho^{AB}=p\ket{\theta^A_1\theta^B_1}\bra{\theta^A_1\theta^B_1}+(1-p)\ket{\theta^A_2\theta^B_2}\bra{\theta^A_2\theta^B_2}\,, \label{rhothetatheta}
\end{align}
$\Phi^{A/B}_\nu$ is taken to be zero here for simplicity, we will see that little changes in the results of our protocol providing the off-diagonal phase on B are all equal, i.e $\{\Phi^B_\nu\}=\Phi_\nu$. $\{\Phi^A_\nu\}$ will be almost entirely irrelevant as we are concerned only with the A-discord. The protocol of course still works if we do not take the set of $\Phi^B_\nu$ to be the same for all $\nu$, and we give an example of this later in section \ref{sec:differentPhis}.

\begin{figure*}
\qquad
\includegraphics[height=.42\textwidth]{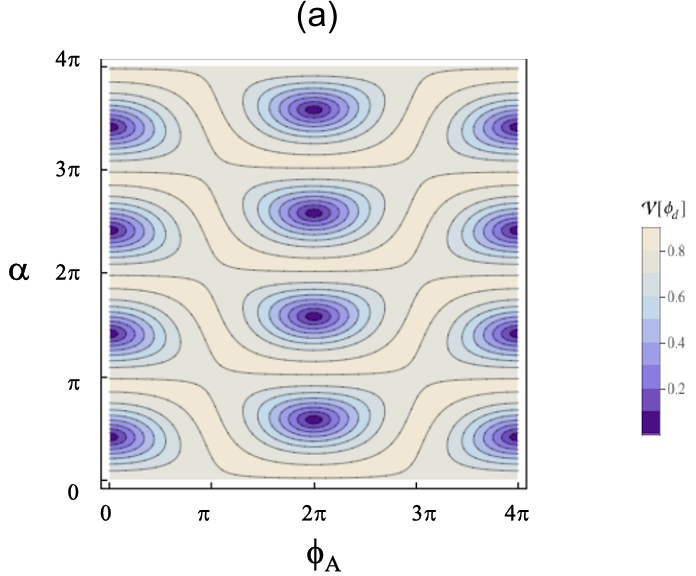}\quad
\includegraphics[height=.42\textwidth]{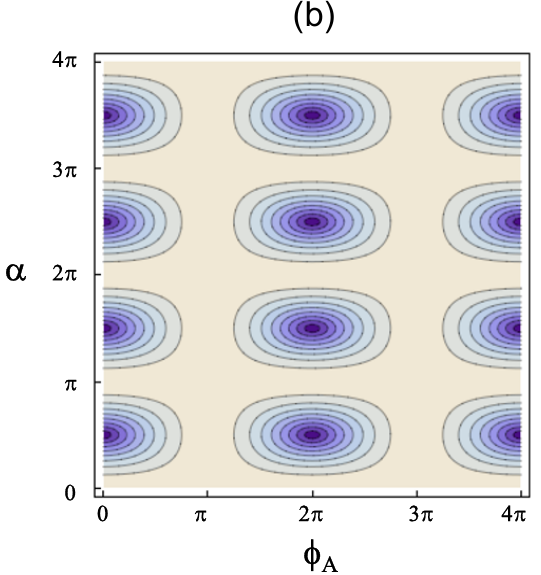}
\vspace*{-6pt}

\caption{ \label{fig5} The visibility landscape, Eq.~\eqref{V}, for the two in-states used in Fig.~\ref{fig2} as a function of parameters $\alpha $ and $\phi _A$ controlling, respectively, the transparency of BS$_1^A$ and the phase difference in   MZI$^A$ (see Fig.~\ref{MZI}). Here we keep fixed the  values of corresponding parameters in MZI$^B$ ($\beta =\pi/3$ and $\phi _B=0$).}
\end{figure*}

According to our scheme we now must extract the value of either $\beta_0$ or $\phi_{B0}$ for a fixed $\alpha$. First, we draw the $\alpha ({\phi _A})$ dependence for the state used as the example in the main text (see Fig.~\eqref{fig5}).

Next, we  arbitrarily choose the value of $\alpha=\pi/3$ and then plot  the visibility as a function of $\beta$ and $\phi_B$.

\begin{figure*}
\includegraphics[height=.45\textwidth]{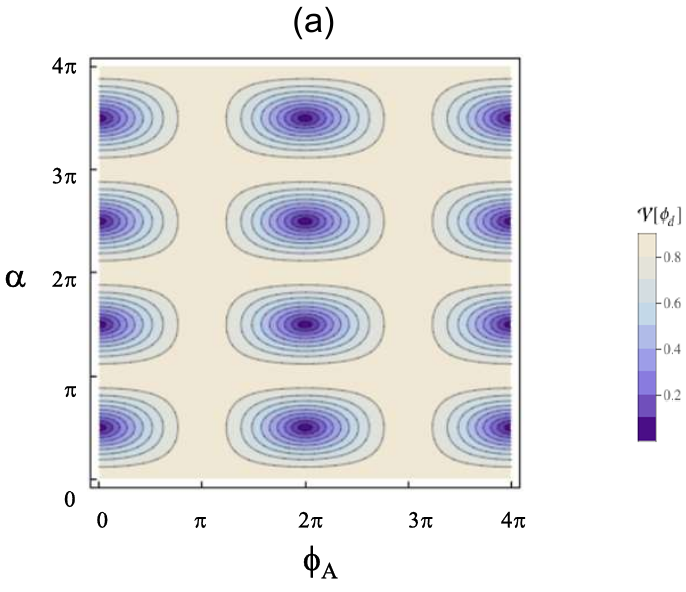}\quad
\includegraphics[height=.4\textwidth]{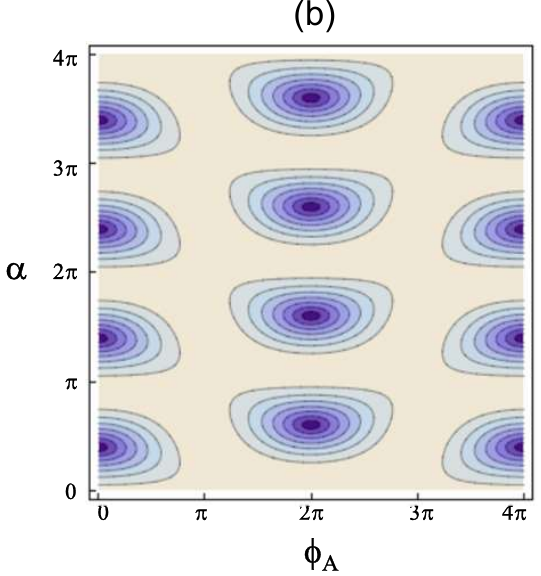}
\vspace*{-6pt}

\caption{Visibility as a function of the phase difference of the first loop $(\phi_B$) against the varying of $BS^B_1$ (parametrised by $\beta$) with $\alpha$ fixed to $\pi/3$. Left: $\rho^{AB}=1/5\big(\ket{++}\bra{++}+4\ket{--}\bra{--}\big)$ Right:$\rho^{AB}=1/2(\ket{\uparrow\uparrow}\bra{\uparrow\uparrow}+\ket{\downarrow+}\bra{\downarrow+})$.} \label{fig:PhiVsBetaQQ}
\end{figure*}
The value of $(\alpha_0,\phi_{A0})$ are given by the coordinates where the visibility drops to zero, we only need one of the components, we pick $\phi_{A0}$ and extract this to be $\phi_{A0}=0,\pi$ in both cases. We may therefore choose either $\phi_{A0}=0,\pi$ for the next step. Actually this would be the correct choice for any state within family we have choosen as our examples (given by Eq.~(\ref{rhothetatheta})). The condition $\{\Phi_{\nu}\}=0$ reduce to the conditions given by,
\begin{align}
\tan(\alpha_0)=\pm\frac{\sum_{\nu=1}^{n} w^{B}_{\nu} \sin(\theta_{\nu})}{\sum_{\nu=1}^{n} w^{B}_{\nu} \cos(\theta_{\nu})}, \hspace{2.5cm} \phi_{A_0}=0,\pi
\end{align}
where $+,-$ solutions correspond to when $\phi_{A_0}=0,\pi$ respectively. If we limit ourselves to states which fall within the family of states where $\{\Phi_{\nu}\}=0$, then step one of our scheme can be skipped and we can proceed with step two by choosing $\phi_B=0,\pi$. Similarly, if $\{\Phi_{\nu}\}=\Phi$ then it can be shown $\phi_{A_0}=\Phi,\Phi+\pi$ in which case we may take $\phi_A=\phi_{A0}$ and skip step 1.

As of yet we can still make no statements about discord, we therefore proceed with step 2 of the protocol; we fix $\phi_{B0}=0$ ($\pi$ would have been an equally appropriate choice for our choosen states) and plot the visibility as a function of $\alpha,\beta$. This will allow us to extract the equation for $\beta_0$,

\begin{figure*}[h!]
\includegraphics[height=.45\textwidth]{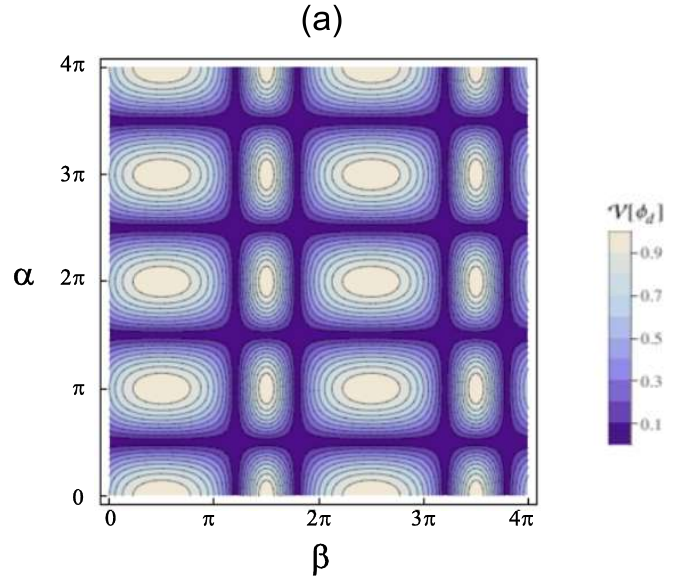}\quad
\includegraphics[height=.45\textwidth]{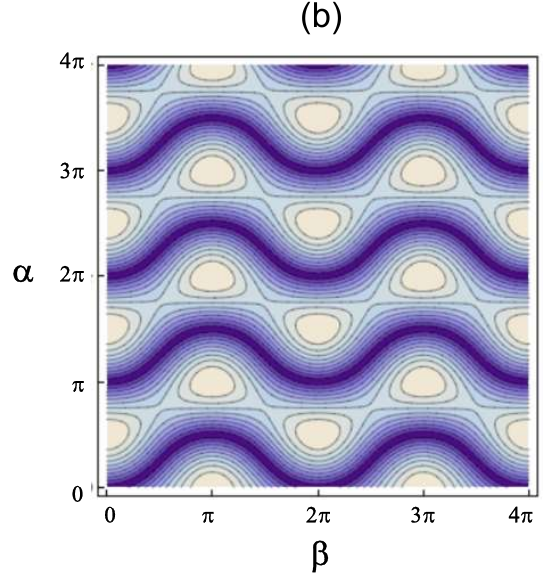}
\vspace*{-6pt}

\caption{Visibility as a function of $\alpha$ and $\beta$ with $\phi_A=\phi_{A_0}$. (a): $\rho^{AB}=1/5\ket{++}\bra{++}+4/5\ket{--}\bra{--}$ (b): $\rho^{AB}=1/2\ket{\uparrow\uparrow}\bra{\uparrow\uparrow}+1/2\ket{+\downarrow}\bra{+\downarrow}$.} \label{fig:AlphaVsBetaQQ}
\end{figure*}
The graphs clearly display what we expect it to, in the case of zero discord state (left) there are straight horizontal lines of zero visibility corresponding to $\beta_0=constant$ (note these grid-type graph characteristic of a classically correlated density matrix). Whilst in the second case the lines of zero visibility are clearly $\alpha$ dependent.

\begin{figure*}[t]

 (a) \hspace*{7.5cm} (b) \vspace*{6pt}

\includegraphics[width=.48\textwidth]{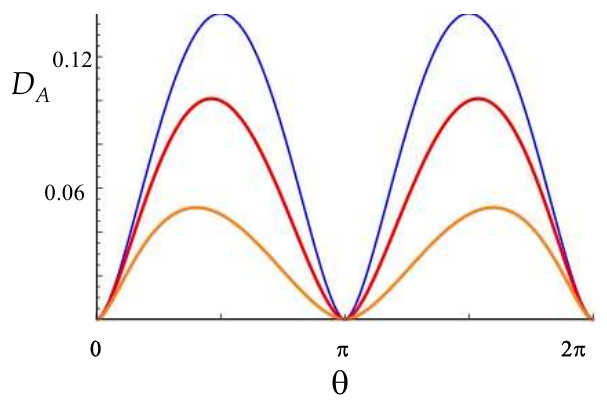}\quad
\includegraphics[width=.48\textwidth]{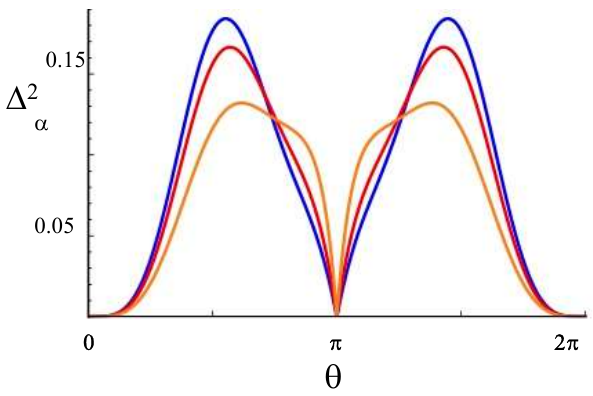}
\vspace*{-6pt}

\caption[Plots of discord and the $\Delta^2_A$-measure for asymmetric graphs.]{  Discord (a) and our quantifier $\Delta^2_\alpha$ (b) for state Eq.~(\ref{thetaArho}) with, $\theta^A=\pi$ (blue), $\theta^A=\pi/2$ (red), $\theta^A=\pi/4$ (orange). }\label{fig:CompareDiscord}
\end{figure*}

The A-discord is sensitive to the states of B-subsystem. We can compare our measure with the discord for a range of density matrices,
\begin{align}
\rho^{AB}=1/2\ket{\uparrow\uparrow}\bra{\uparrow\uparrow}+1/2\ket{\theta^A\theta}\bra{\theta^A\theta}\,,\label{thetaArho}
\end{align}
by changing $\theta^B$ in the state given by Eq.~ (\ref{thetaArho}) and compare it to $\Delta^2_\alpha$.
Discord is plotted for different values of $\theta^A$ in Fig. (\ref{fig:CompareDiscord}) and the complimentary values of $\Delta^2_\alpha$ are given in Fig. \ref{fig:CompareDiscord}. We see that, for discord, the peaks of the curves shift and the amplitude is dimished as $\theta^B\rightarrow 0$. The peaks of $\Delta^2_\alpha$ also decrease as $\theta^A$ is reduced, though the position of the $\theta$-dependent peaks of each function match Fig. (\ref{fig:CompareDiscord}) less well as $\theta^A$ decreases. This measure $\emph{always}$ matches the zeros of discord providing $\{\Phi_{A_\nu}\}$$=\Phi_A$ however, and thus never produces a false witness in this case.

For $\{\Phi^A_\nu\}\neq \Phi_A$ the measure $\Delta^2_\alpha>0$ can fail as a necessary condition for discord some specific states (inspite of the fact our protocol still provides a necessary condition, i.e there will still be no $\beta$-independent lines of zero visibility in the $\alpha-\beta$ or $\beta-\phi_A$ visibility landscapes ). Therefore if we do not fix  $\{\Phi^A_\nu\}$ to a non-zero $\Delta^2_\alpha$ becomes only a sufficient condition for discord. This is a result of some select choices of discorded $\rho^{AB}$ having a $\alpha_0$ which is not $\beta$ dependent, but a diagonalising phase, $\phi_{A_0}$, which is. In such a situation we may consider a similar function to $f_\alpha(\beta)$ but based on $\phi_{A_0}$ as opposed to $\alpha_0$ we consider,
 \begin{align}
    f_\phi(\beta)\equiv\cos^2(\phi_{A_0})=\mathcal{N}_x^2/(\mathcal{N}_x^2+\mathcal{N}_y^2)
 \end{align}
 We may then consider the standard deviation of this quantity $\Delta_\phi^2$ (defined similarly to Eq. (\ref{W})) which, when it is $>0$, also individually provides a sufficient condition for discord. Together with $\Delta^2_\alpha$, however, it provides a necessary condition, i.e $\Delta_\phi^2+\Delta^2_\alpha>0$ is a necessary condition for discord. We give an example where $\Delta^2_\alpha$ on it own fails as a witness in the next section, but see even with being the case that it is clear from the $\alpha-\beta$-visiblity landscape whether the state is discorded or not.

\begin{figure*}
\includegraphics[height=.45\textwidth]{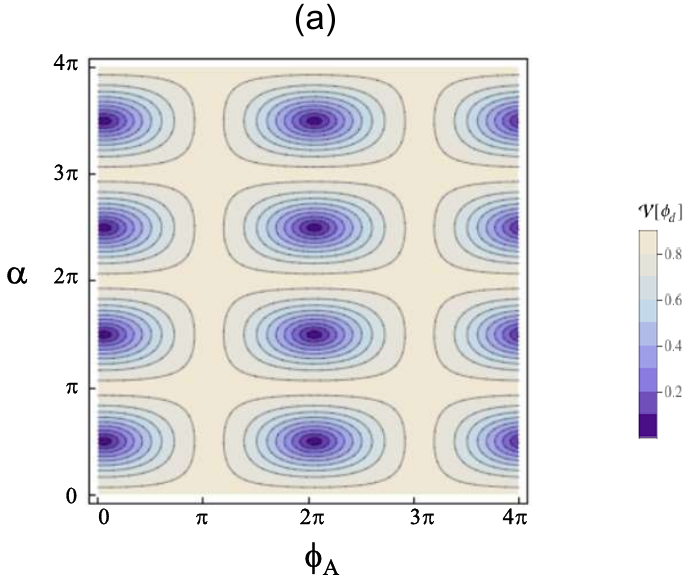}\quad
\includegraphics[height=.45\textwidth]{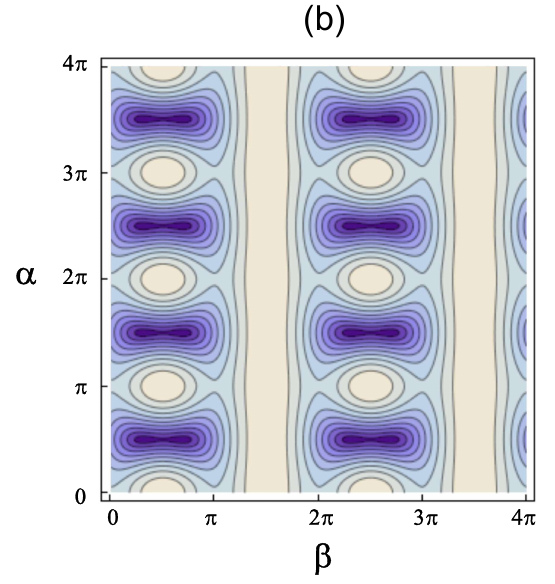}
\vspace*{-6pt}

\caption{  The visibility plots for the density matrix given by Eq. (\ref{rhophi} with $\Phi_1=0$ and $\Phi_2=\pi/2$, (a) Shows visibility as function of $\alpha,\phi_A$ with $\beta=2\pi/3$, (b) gives visibility plot as function of $\alpha,\beta$ now fixing $\phi_A=\arctan\big(\frac{2-\sqrt{3}}{3}\big)$ (see Eq. (\ref{phiB0differentphis}).}\label{fig:differentphis}
\end{figure*}

\textbf{\textit{States with phase differences}.} \label{sec:differentPhis}
Up until this point we only considered examples of our protocol for when density matrices $\rho^A_\nu$ had the same off-diagonal phase, but our protocol also works for states where the phases are different. Below we will give an example of what we would expect using our protocol for a state with off-diagonal phase on A, we will continue to assume there is no off-diagonal phase on $B$ for the sake of simplicity. We consider the state,
\begin{align}
\rho^{AB}=\frac{1}{2}\ket{+;+,\Phi_1}\bra{+;+,\Phi_1}+\frac{1}{2}\ket{-;-,\Phi_2}\bra{-;-,\Phi_2} \label{rhophi}
\end{align}
where $\ket{\theta_{A_\nu};\theta_{B_\nu},\Phi_{B_\nu}}\equiv\ket{\theta_{A_\nu}}\otimes \ket{\theta_{B_\nu},\Phi_{B_\nu}}\equiv \ket{A_\nu}\big|_{\Phi_{A_\nu}=0}\otimes \ket{B_\nu}$. The above state has zero discord providing that $\Phi_1=\Phi_2+n\pi$ where n is an integer. We will arbitrarily pick $\Phi_1=0$ and $\Phi_2=\pi/2$ so that the resulting state is discorded and then proceed to check this using our protocol. First we plot the visibility graph with fixed $\beta$ (we choose $\beta=2\pi/3$) as function of $\alpha$ and $\phi_A$, this is shown in Fig. (\ref{fig:differentphis}a).
These graphs appear similar to the ones shown earlier, but the values of $\phi_A$ where visibility drops to zero are now slightly shifted from zero and $\pi$. Their positions are given by,
\begin{align}
\phi_{A0}=\arctan\big(\frac{2-\sqrt{3}}{3}\big)\quad{\rm mod}\quad\pi.
\label{phiB0differentphis}
\end{align}
Fixing $\phi_A$ to the first of these values we then plot the visibility as a function of $\alpha$ and $\beta$, this is given in (\ref{fig:differentphis}b).

\begin{figure*}
\includegraphics[width=.4\textwidth]{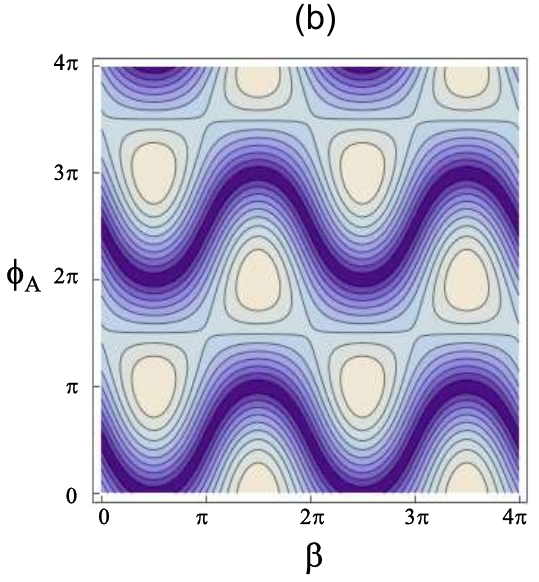}\qquad\qquad
\includegraphics[width=.52\textwidth]{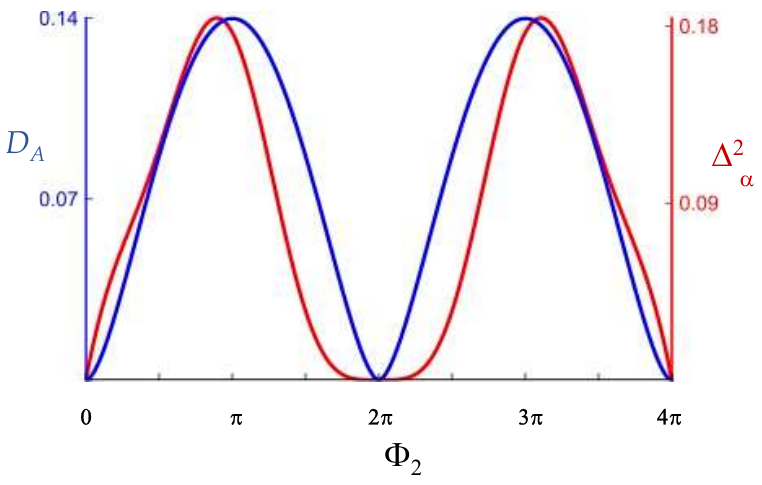}
\vspace*{-6pt}

\caption{(a) Visibility plot for the density matrix given by Eq. (\ref{rhophi} with $\Phi_1=0$ and $\Phi_2=\pi$, plots are given as a function of $\phi_A$ and $\alpha$ with $\beta=\beta_0=\pi/2$. (b) demonstrates discord and $\Delta^2_\phi$ for the state described by Eq. (\ref{rhophi} for a range of $\Phi_2$ with $\Phi_1=0$.}\label{fig:differentphis_discord}
\end{figure*}
The difference between the state shown here and those given previously is now apparent. It is even more stark here that the parameters $\alpha_0,\phi_{A0}$ depend on $\beta$. This is due to the fact the value of $\phi_{A0}$ is now also $\beta$-dependent (in previous examples where $\{\Phi_\nu\}=\Phi$ it was only $\alpha_0$ which depended on $\beta$). This means that as $\beta$ is changed we no longer have the correct value of $\phi_{A}$ which diagonalises the state, and since to diagonalise the state both $\alpha=\alpha_0$ and $\phi_A=\phi_{A0}$ must be true there are large regions of Fig. \ref{fig:differentphis_discord}(b) where the state can not be diagonalised and therefore visibility can not go to zero. A very clear demonstration that the state is discorded. This is an example of where if we take $\Delta^2_\alpha$ alone we would not be able to tell the state were discorded (in spite of how obvious it is from the Fig. \ref{fig:differentphis}(b)), $\Delta_\phi^2$ however behaves similarly to discord as seen in Fig. \ref{fig:differentphis_discord}(b). $\Delta_\phi^2$ may be extracted from visibility landscapes like Fig. $\ref{fig:differentphis_discord}(a)$ in which $\alpha$ is fixed to $\alpha_0$ and we have plotted the visibility as a function of $\phi_A$ and $\beta$.

\begin{figure*}[h!]
\begin{minipage}{0.45\textwidth}\includegraphics[width=\textwidth]{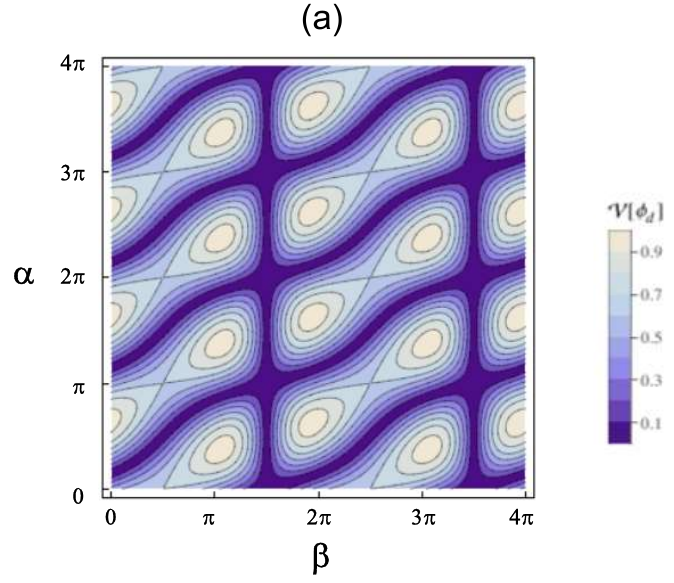}\end{minipage}
\begin{minipage}{0.45\textwidth}\includegraphics[width=\textwidth]{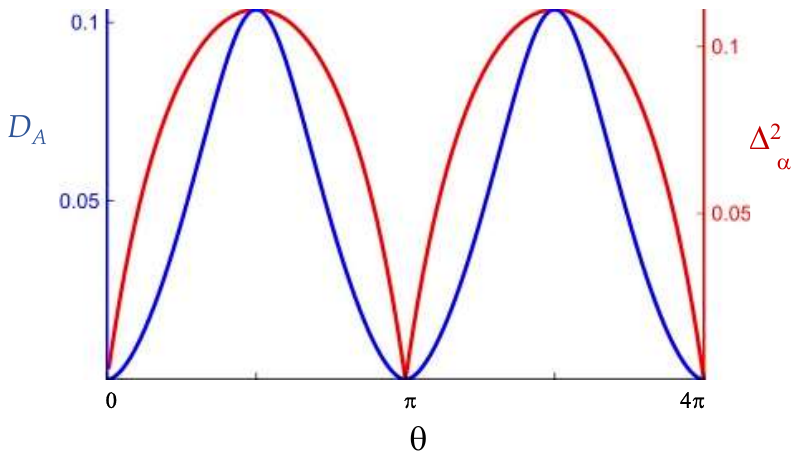}\end{minipage}
\vspace*{-6pt}
\caption{(a) Shows a visibility plot of density matrix given by Eq. (\ref{eq:three state}) with $\theta=\pi/2$, the tell-tale signs of discord are present in the curviness of the zero-visibility lines. (b) plots the discord of Eq. (\ref{eq:three state}) for different $\theta$ and contrasts it to the measure of curviness which can be extracted from graphs like that shown in (a).} \label{fig:3states}
\end{figure*}

\textbf{\textit{Example with three states}.}
We have previously limited ourselves to examples with $n=2$ in a separable density matrix described by Eq. (\ref{sepstate}), we briefly consider an example with $n=3$,
\begin{align}
\rho^{AB}=1/3\ket{\uparrow\uparrow}\bra{\uparrow\uparrow}+1/3\ket{\downarrow\downarrow}\bra{\downarrow\downarrow}+1/3\ket{\theta\theta}\bra{\theta\theta} \label{eq:three state}
\end{align}
Since the state is real we know that the correct choice of the diagonalising rotation is $\phi_{A_0}=0$, fixing $\phi_A$ to this value we can then plot the visibility as a function $\alpha$ and $\beta$, Fig (\ref{fig:3states}a) gives a snapshot of one of these visibility graphs for $\theta=\pi/2$. We see the characteristic waviness which correctly tells us the state is discorded, $\Delta_\alpha^2$ for this value of  $\theta$ can be extracted from this graph. We plot $\Delta_\phi^2$ for different $\theta$ and contrast it to A-discord in Fig. (\ref{fig:3states}b). We see that once again the most important features of discord are mirrored in $\Delta^2_\phi$.

\textbf{\textit{Experimentally obtaining measure for 3D lines of zero visibility}.}
In order to experimentally obtain $\Delta^2_\alpha+\Delta_{\phi}^2$ for a completely general state, one considers the full zero-visibility lines in three-dimensional space $(\alpha,\,\beta,\,\phi_A)$. The discord quantifier is extracted from  this line by calculating $\Delta^2_\alpha+\Delta_{\phi}^20$, which is zero only if  discord is absent. The measures $\Delta^2_\alpha$  and $\Delta^2_\phi$ can be obtained by the projection of the line onto the $\alpha-\beta$ and $\phi_A-\beta$ planes respectively.
\vspace*{24pt}
 \begin{center}
    \underline{{\quad\quad\quad}}
 \end{center}

\vspace*{12pt}

[1] H. Ollivier and W. H. Zurek, Phys. Rev. Lett. \textbf{88}, 017901 (2001).

[2] L. Henderson and V. Vedral, J. Phys. A \textbf{34}, 6899 (2001).

[3] V. Vedral, arXiv:quant-ph , 1702.01327 (2017).

[4] B. Daki\'c, V. Vedral, and
C. Brukner, Phys. Rev. Lett. \textbf{105}, 190502 (2010).

[5] A. Ferraro, L. Aolita, D. Cavalcanti, F. M. Cucchietti, and A. Ac\'{i}n, Phys. Rev. A \textbf{81}, 052318 (2010).

[6] K. Modi, A. Brodutch, H. Cable, T. Paterek, and V. Vedral, Rev. Mod. Phys. \textbf{84}, 1655 (2012).
\end{document}